\documentclass{article}
\usepackage{spconf,amsmath,amssymb,graphicx,booktabs,float,cite,hyperref}

\title{SCIC: SCOPE- AND CODEBOOK-AWARE INSTRUCTION\\CONDITIONING FOR SPEAKER-ADAPTED EXPRESSIVE TTS}
\name{Longyu Lu, Zongwei Du, Mengtao Xing, Zhuoqun Liu, Zifan Guan, Meiguang Jin\textsuperscript{*}\thanks{*Corresponding author.}, Junfeng Ma}
\address{TaoLive-AIGC Team \\
Taobao \& Tmall Group of Alibaba}

\begin{document}
\ninept
\maketitle

\begin{abstract}
Long-form live-streaming TTS requires context-dependent prosody and paragraph-level coherence. However, many existing instruction-based TTS systems use global or uniform conditions, providing limited explicit control over clause-level relative prosodic changes. We introduce Speaker-Relative Inline Prosody Control, where each Pitch, Energy, or Speed instruction targets a clause relative to the preceding clause from the same speaker, while Pause uses an absolute duration interval. In codec-based TTS, Speed and Pause affect sequence length, whereas Pitch and Energy rely on residual codebooks. By analyzing Qwen3-TTS RVQ codebooks, we find that Energy concentrates in early residual codebooks, whereas Pitch accumulates across a deeper prefix. We therefore propose Scope- and Codebook-Aware Instruction Conditioning (SCIC), combining a Temporal Instruction Router for frame-level tag activation with Tag-Specific Codebook Weighting over residual codebooks. SCIC improves speaker-relative Pitch and Energy control over standard instruction fine-tuning using text-token tags. We further apply multi-reward GDPO post-training to jointly optimize control and quality, improving control accuracy while preserving CER and speaker similarity. In long-form synthesis, SCIC produces a more distinct paragraph-level expressive hierarchy than speaker-adapted SFT without instructions. Audio demos are available at: \url{https://taoliveaigc.github.io/SCIC/}.
\end{abstract}


\section{Introduction}
\label{sec:intro}

In long-form live-streaming scenarios, speech exhibits local variations in pitch, energy, speaking rate, and pauses. TTS systems for this setting require fine-grained instruction control while maintaining paragraph-level coherence. Recent systems support natural-language instructions \cite{instructtts,prompttts,prompttts2,promptstyle,controlspeech,flespeech,ovinstructtts,flexivoice,bagpipertts,cosyvoice3,hdppt}, categorical, continuous, or relative attribute control \cite{sparktts,voxtream2,restyle}, and fine-grained local control through inline or hierarchical annotations \cite{magictts,wordvoice,qwenaudio3tts,fishaudios2,any2speech,emotratts}. Preference or policy optimization has also improved instruction adherence and controllability \cite{flexivoice,fishaudios2}. However, existing methods do not jointly formulate inline prosody as a clause-level change relative to the preceding same-speaker clause and model its temporal scope and RVQ codebook allocation. This gap limits coherent prosodic progression in long-form synthesis.

We address this problem through Speaker-Relative Inline Prosody Control. An inline tag placed before a target clause specifies a local Pitch, Energy, Speed, or Pause change. For Pitch, Energy, and Speed, the preceding clause from the same speaker serves as the reference anchor for the controlled clause; Pause is specified by an absolute duration interval. This formulation avoids imposing one absolute acoustic target across speakers and instead preserves speaker-specific prosodic variation. It also provides a direct interface for constructing paragraph-level expressive progression from a sequence of locally controlled clauses. Our setting is speaker-adapted rather than zero-shot speaker generalization: all controlled systems are adapted to the target voices during supervised fine-tuning and evaluated on held-out utterances from those speakers.

A remaining challenge is that different prosodic attributes are realized through different parts of a codec-based autoregressive generator. In Qwen3-TTS \cite{qwen3tts}, the Main Talker predicts the first codebook $q_0$ and advances the acoustic-frame sequence, while Multi-Token Prediction (MTP) predicts residual codebooks $q_1$--$q_{15}$ within each frame. Speed and Pause primarily affect temporal progression and sequence length and can therefore be controlled through the existing $q_0$ path. Pitch and Energy, by contrast, depend on acoustic information distributed across the residual codebooks. Standard instruction fine-tuning represents inline tags only as text tokens and does not explicitly determine either when a Pitch/Energy tag should be active or how strongly it should affect each residual codebook.

To identify the relevant codec structure, we analyze Qwen3-TTS through RVQ Codebook Diagnosis, exchanging projected codebook contributions between original and attribute-transformed speech. The resulting cumulative-transfer trajectories differ substantially by attribute: Energy effects saturate within the first few residual codebooks, whereas Pitch transfer accumulates across a substantially deeper codebook prefix. Motivated by this observation, we propose Scope- and Codebook-Aware Instruction Conditioning (SCIC). Its Temporal Instruction Router predicts frame-level activation for each inline Pitch/Energy tag, and its Tag-Specific Codebook Weighting modulates the signed conditioning strength across $q_1$--$q_{15}$. The two components jointly inject a localized SCIC conditioning signal into MTP, while Speed and Pause remain on the autoregressive $q_0$ path. We further post-train SCIC-SFT with multi-reward GDPO to jointly optimize prosody control, Pause accuracy, intelligibility, and speaker similarity. Experiments confirm improved local control while preserving synthesis quality and long-form coherence.

Our key contributions are as follows. First, we formulate Speaker-Relative Inline Prosody Control with absolute Pause intervals and relative Pitch, Energy, and Speed Up/Down instructions. Second, RVQ Codebook Diagnosis reveals attribute-dependent behavior: Energy transfer saturates within the first few residual codebooks, whereas Pitch transfer accumulates across a substantially deeper codebook prefix. Third, we propose SCIC, which combines a Temporal Instruction Router with Tag-Specific Codebook Weighting, and further improve SCIC-SFT through multi-reward GDPO post-training with explicit control and quality objectives.

\section{Methodology}
\label{sec:method}

\begin{figure*}[t]
\centering
\includegraphics[width=0.96\textwidth]{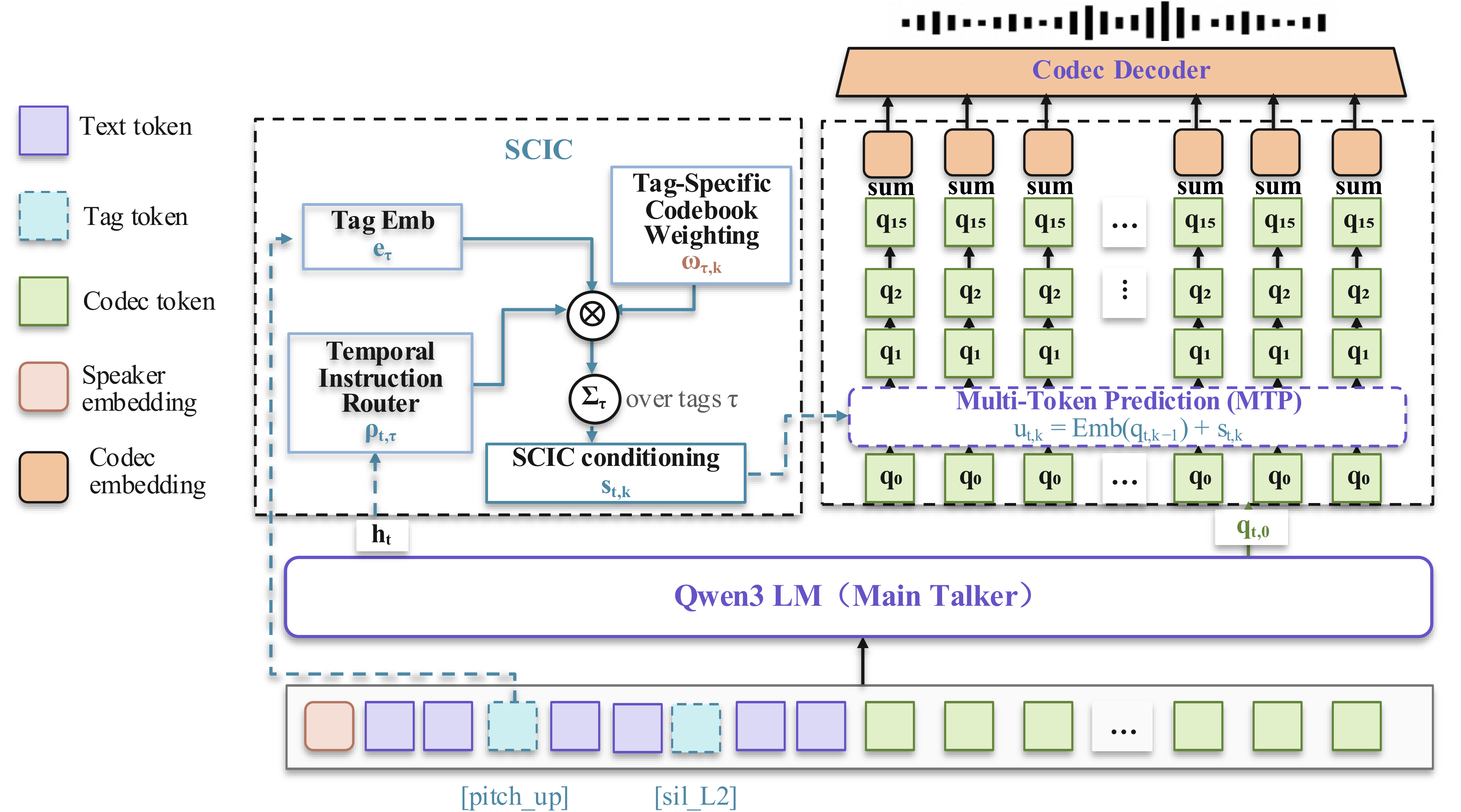}
\vspace{-5pt}
\caption{SCIC in Qwen3-TTS. The Main Talker predicts $q_0$ and provides $h_t$. For each Pitch/Energy tag, its tag embedding is scaled by the Temporal Instruction Router output and the tag-specific codebook weight to form $s_{t,k}$, which conditions MTP prediction of $q_1$--$q_{15}$.}
\label{fig:architecture}
\end{figure*}

\begin{figure*}[t]
\centering
\includegraphics[width=0.92\textwidth]{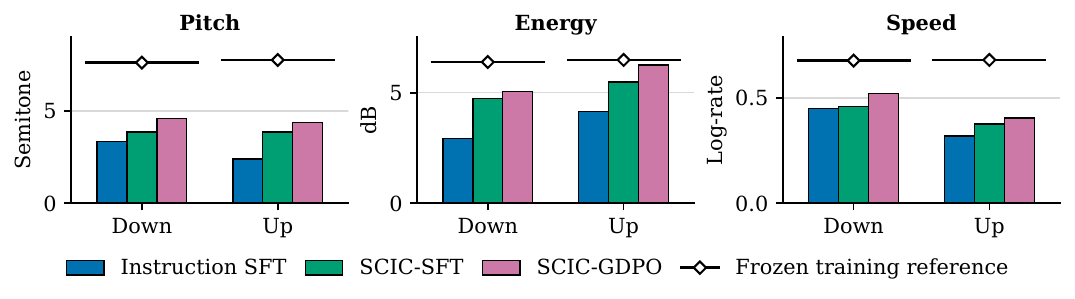}
\caption{Median speaker-relative change magnitudes after direction normalization. Down values are sign-flipped so positive values indicate the requested direction; black lines and diamonds denote frozen training-reference magnitudes.}
\label{fig:amplitude}
\end{figure*}

\subsection{Overall framework}

Figure~\ref{fig:architecture} illustrates the generation pipeline. The Main Talker predicts $q_0$ and provides the frame-level state $h_t$ to SCIC. For each Pitch/Energy tag $\tau$, its tag embedding $e_\tau$, presence indicator $\chi_\tau$, Router output $\rho_{t,\tau}$, and codebook weight $\omega_{\tau,k}$ form $s_{t,k}$ after summation over tags. At residual step $k$, this signal is added to the preceding codebook-token embedding, yielding $u_{t,k}=E_{k-1}(q_{t,k-1})+s_{t,k}$. MTP uses $h_t$ and the conditioned inputs to predict $q_1$--$q_{15}$, whose projected contributions are summed and decoded. Multi-reward GDPO further optimizes control, intelligibility, and speaker similarity.

\subsection{Speaker-relative inline control formulation}

We first formalize how inline instructions specify local prosodic targets. Each instruction associates a target clause with either a speaker-relative change in Pitch, Energy, or Speed, measured against the immediately preceding clause from the same speaker, or an absolute Pause interval at the instructed boundary. Let $c_{i-1}$ and $c_i$ denote the reference and target clauses, respectively, with the inline tag placed immediately before $c_i$. For voiced-median pitch $F_i$, shared-voiced Top-80\% RMS energy $E_i$, and aligned-character speaking rate $r_i$, the adjacent-clause changes are
\begin{align}
\Delta_i^{\rm pit}&=12\log_2(F_i/F_{i-1}), &
\Delta_i^{\rm eng}&=20\log_{10}(E_i/E_{i-1}),\nonumber\\
\Delta_i^{\rm spd}&=\log(r_i/r_{i-1}),&
 m_i^a&=\eta_i\Delta_i^a,
\label{eq:measure}
\end{align}
where $a\in\{\mathrm{pit},\mathrm{eng},\mathrm{spd}\}$ indexes the controlled attribute, and $\eta_i=+1$ for an Up instruction and $-1$ for a Down instruction. Thus $m_i^a$ is the direction-normalized change, with a positive value always indicating movement in the requested direction. The logarithmic definitions provide symmetric representations of reciprocal Up and Down changes and avoid imposing one absolute acoustic target across speakers.

Pause is treated as an absolute-duration task with three reported levels: L1 $[0.50,0.75]$, L2 $[0.75,1.00]$, and L3 $[1.00,1.25]$ seconds.

\subsection{RVQ codebook diagnosis}

Given an original waveform $x$, we construct a duration-matched counterpart $x'$ by applying either an upward or downward Pitch shift, or an increase or decrease in Energy. Codec encoding produces $C,C'\in\mathbb{Z}^{T\times16}$, where $T$ is the number of codec frames and column $k$ contains codebook-$k$ tokens. Let $\ell_k$ and $\ell'_k$ denote the projected contribution of codebook $k$ to the decoder input, giving $z=\sum_{k=0}^{15}\ell_k$ and $z'=\sum_{k=0}^{15}\ell'_k$. We exchange these projected contributions rather than discrete token columns. For a codebook set $S\subseteq\{0,\ldots,15\}$,
\begin{align}
\delta_S&=\sum_{k\in S}(\ell'_k-\ell_k),&
\widetilde z_S^{\rightarrow}&=z+\delta_S,&
\widetilde z_S^{\leftarrow}&=z'-\delta_S.
\label{eq:swap}
\end{align}
Here $\delta_S$ is the transformed-minus-original contribution from $S$; $\widetilde z_S^{\rightarrow}$ inserts it into $z$, whereas $\widetilde z_S^{\leftarrow}$ removes it from $z'$. Let $D$ denote codec decoding and $\phi_a$ the measurement of attribute $a\in\{\mathrm{pit},\mathrm{eng}\}$. Transfer relative to the complete codec-reconstructed change is
\begin{align}
\mathrm{Tr}_{S,a}^{\rightarrow}&=
\frac{\phi_a(D(\widetilde z_S^{\rightarrow}))-\phi_a(D(z))}
{\phi_a(D(z'))-\phi_a(D(z))},\nonumber\\
\mathrm{Tr}_{S,a}^{\leftarrow}&=
\frac{\phi_a(D(z'))-\phi_a(D(\widetilde z_S^{\leftarrow}))}
{\phi_a(D(z'))-\phi_a(D(z))}.
\label{eq:transfer}
\end{align}
Forward intervention inserts the transformed contributions into the original reconstruction, while reverse intervention removes them from the transformed reconstruction. Negative transfer indicates movement opposite to the transformation, and values above 100\% indicate overshoot. These ratios describe attribute transfer under latent intervention rather than additive information percentages.

Energy rapidly saturates within $q_1$--$q_3$, whereas Pitch requires a deeper interacting prefix and approaches saturation around $q_1$--$q_{10}$. These different trajectories motivate tag-specific codebook weights rather than uniform injection or one fixed profile.

\subsection{Scope- and codebook-aware instruction conditioning}

SCIC factorizes local instruction conditioning along the temporal and codebook axes. Let $\tau$ index the four inline Pitch/Energy instruction-tag types, $\chi_\tau\in\{0,1\}$ indicate whether tag $\tau$ occurs in the input, and $e_\tau=P_{\rm text}\mathrm{Emb}_{\rm text}(\tau)$ be its projected text representation. Here $\mathrm{Emb}_{\rm text}(\tau)$ is the embedding of tag token $\tau$, and $P_{\rm text}$ is the existing Qwen3-TTS text projection that maps it to the hidden dimension shared by the Main Talker and MTP inputs. Given the Main Talker state $h_t$ associated with codec frame $t$, the Temporal Instruction Router predicts
\begin{equation}
\rho_{t,\tau}=\sigma\!\left([W\,\mathrm{LN}(h_t)+b]_\tau\right)\chi_\tau.
\label{eq:router}
\end{equation}
Here $W$ and $b$ are learned Router parameters, $\mathrm{LN}$ is layer normalization, and $\sigma$ is the logistic sigmoid. Thus $\rho_{t,\tau}\in[0,1]$ specifies the frame-level activation strength of tag $\tau$; multiplication by $\chi_\tau$ makes absent tags exactly inactive. Aligned clause spans provide binary frame-level supervision. The Router loss balances positive and negative frames separately for each supervised tag type so that long inactive regions do not dominate training.

For residual codebook $k\in\{1,\ldots,15\}$, Tag-Specific Codebook Weighting assigns each tag a zero-initialized parameter $\omega_{\tau,k}$. The localized conditioning signal and conditioned input are
\begin{align}
s_{t,k}&=\sum_\tau \rho_{t,\tau}\omega_{\tau,k}e_\tau,\nonumber\\
u_{t,k}&=\mathrm{Emb}_{k-1}(q_{t,k-1})+s_{t,k}.
\label{eq:inject}
\end{align}
MTP predicts $(\hat q_{t,1},\ldots,\hat q_{t,15})$ from $h_t$ and $(u_{t,1},\ldots,u_{t,15})$. Here $\mathrm{Emb}_{k-1}$ is the codec-token embedding used when predicting $q_{t,k}$. The Router controls temporal scope, while $\omega_{\tau,k}$ controls signed strength across $q_1$--$q_{15}$. Pitch and Energy use this explicit SCIC conditioning path. Training combines generation and balanced frame-level Router losses, $\mathcal{L}=\mathcal{L}_{\rm gen}+\lambda_r\mathcal{L}_{\rm router}$, where $\lambda_r$ weights the Router objective.

\subsection{Multi-reward GDPO post-training}

SCIC-SFT is post-trained with bounded control and quality rewards. For Pitch, Energy, or Speed, let $m$ be the direction-normalized change, $T=T_{v,a}=1.8\sigma_{v,a}$ the frozen speaker-specific threshold, and $P=P_{v,a,d}$ the direction-specific training-set 99th-percentile cap, with $0<T<P$. The control reward is
\begin{equation}
R_{\rm ctrl}(m)=
\begin{cases}
\max(-1,m/T), & m\leq0,\\
m/T, & 0<m<T,\\
1, & T\leq m\leq P,\\
\max\!\left(-1,1-\dfrac{2(m-P)}{P-T}\right), & m>P.
\end{cases}
\label{eq:control_reward}
\end{equation}
This rewards requested-direction changes above $T$ and penalizes incorrect or excessive responses. For pause duration $p$ and target interval $[l,h]$, let $\operatorname{dist}(p,[l,h])=\max(l-p,0,p-h)$. The Pause reward is
\begin{equation}
R_{\rm pau}(p)=
\begin{cases}
1,&p\in[l,h],\\
\max\!\left(-1,1-\dfrac{2\operatorname{dist}(p,[l,h])}{h-l}\right),&\text{otherwise}.
\end{cases}
\label{eq:pause_reward}
\end{equation}
Bounded CER and speaker-similarity rewards preserve intelligibility and speaker identity. Following GDPO \cite{gdpo}, each objective $o$ is standardized across the $G$ responses $j$ to input $i$ before weighted aggregation:
\begin{align}
A_o^{(i,j)}&=\frac{R_o^{(i,j)}-\mu_{i,o}}{\sigma_{i,o}+\epsilon},&
A^{(i,j)}&=\sum_o w_o A_o^{(i,j)}.
\label{eq:gdpo}
\end{align}
Here $\epsilon$ ensures numerical stability and $w_o$ weights objective $o$, preventing reward-scale differences from dominating optimization.

\begin{table*}[t]
\caption{RVQ transfer (\%). Left: selected single-codebook results with speaker-bootstrap 95\% confidence intervals. Right: cumulative transfer over residual-codebook prefixes.}
\label{tab:rvq}
\centering
\small
\begin{tabular}{@{}c@{\hspace{1.5em}}c@{}}
\textbf{Single-codebook transfer} & \textbf{Cumulative transfer} \\[2pt]
{\setlength{\tabcolsep}{1.7pt}
\begin{tabular}{crrrr}
\toprule
Codebook & Pitch Fwd. & Pitch Rev. & Energy Fwd. & Energy Rev. \\
\midrule
$q_0$ & 0.0 [-0.0,0.1] & 0.5 [-0.2,1.7] & -0.0 [-0.1,0.0] & -0.0 [-0.1,0.0] \\
$q_1$ & -5.0 [-7.4,-2.8] & 18.3 [15.7,21.0] & \textbf{83.5 [82.7,84.3]} & \textbf{85.3 [84.5,86.1]} \\
$q_2$ & -0.1 [-1.6,1.2] & 12.7 [10.0,16.1] & 6.6 [6.1,7.0] & 8.1 [7.7,8.6] \\
$q_3$ & 3.7 [2.6,4.7] & 10.2 [7.7,12.9] & 2.5 [2.2,2.8] & 3.1 [2.7,3.5] \\
$q_4$ & 2.1 [1.4,2.9] & 10.2 [7.3,13.0] & 1.6 [1.3,2.0] & 1.9 [1.6,2.1] \\
$q_5$ & 0.8 [0.2,1.2] & 2.9 [-0.9,5.5] & 1.0 [0.8,1.2] & 1.2 [0.9,1.5] \\
\bottomrule
\end{tabular}}
&
{\setlength{\tabcolsep}{2.0pt}
\begin{tabular}{llcc}
\toprule
Attribute & Codebooks & Forward & Reverse \\
\midrule
Energy & $q_1$--$q_2$ & 91.0 & 92.1 \\
Energy & $q_1$--$q_3$ & \textbf{94.0} & \textbf{94.7} \\
Pitch & $q_1$--$q_4$ & 52.9 & 68.1 \\
Pitch & $q_1$--$q_6$ & 83.4 & 88.7 \\
Pitch & $q_1$--$q_8$ & 91.0 & 96.5 \\
Pitch & $q_1$--$q_{10}$ & \textbf{95.4} & \textbf{99.6} \\
\bottomrule
\end{tabular}}
\end{tabular}

\vspace{4pt}
\caption{Control and quality results. Pitch, Energy, Speed, Pause ACC, and CER are percentages; SIM is cosine similarity, and mean Pause durations are in seconds.}
\label{tab:main}
\centering
\small
\setlength{\tabcolsep}{2.5pt}
\begin{tabular}{@{}lrrrrrrrrrrrr@{}}
\toprule
Model & Pitch & Energy & Speed & \multicolumn{4}{c}{Pause ACC} & \multicolumn{3}{c}{Mean Pause} & CER$\downarrow$ & SIM$\uparrow$ \\
\cmidrule(lr){5-8}\cmidrule(lr){9-11}
& & & & L1 & L2 & L3 & Macro & L1 & L2 & L3 & & \\
\midrule
Instruction SFT & 82.83 & 85.86 & 90.89 & 92.41 & 92.55 & 88.81 & 91.26 & 0.615 & 0.863 & 1.110 & 1.68 & 0.8721 \\
SCIC-SFT & 90.14 & 96.39 & 91.40 & 94.34 & 92.79 & 88.65 & 91.93 & 0.618 & 0.856 & 1.101 & 1.63 & 0.8719 \\
SCIC-GDPO & \textbf{92.48} & \textbf{97.16} & \textbf{93.86} & \textbf{95.82} & \textbf{95.90} & \textbf{93.68} & \textbf{95.13} & 0.635 & 0.876 & 1.136 & 1.64 & 0.8718 \\
\bottomrule
\end{tabular}
\vspace{-4pt}
\end{table*}

\section{Experiments}
\label{sec:experiments}

\subsection{Experimental setup}

\textbf{Dataset.} Our supervised data comprise approximately 1,000 hours of natural Chinese speech from 30 speakers. Each recording is at most 90 seconds long, and the processed corpus contains 132,018 instruction tag. All evaluation utterances are disjoint from the supervised and post-training data.

\textbf{Implementation details.} We initialize all systems from a live-streaming Qwen3-TTS CPT checkpoint and then perform supervised fine-tuning on our instruction data. Instruction SFT uses the same inline tags and supervised data but represents the instructions only through text-token conditioning. SCIC-SFT additionally introduces the Temporal Instruction Router and Tag-Specific Codebook Weighting, and SCIC-GDPO further applies multi-reward post-training. Supervised fine-tuning uses rank-32 LoRA \cite{lora} with $\alpha=64$ on attention and MLP projections. AdamW uses learning rates of $10^{-5}$ for LoRA and $10^{-4}$ for control embeddings, the Router, and tag-specific codebook weights. We train on eight NVIDIA RTX PRO 5000 72GB GPUs with a per-GPU batch size of 8. SCIC-GDPO samples eight responses per input and uses a learning rate of $5\times10^{-6}$.

\textbf{Evaluation metrics.} We evaluate 1,800 Pitch/Energy/Speed samples using control-direction accuracy and median direction-normalized change, and 1,200 samples per Pause level using interval-hit accuracy. Qwen3-ForcedAligner-0.6B \cite{qwen3asr} provides alignment, Qwen3-ASR \cite{qwen3asr} measures CER on 1,800 texts, and WeSpeaker \cite{wespeaker} measures similarity on 1,200 samples. Router localization is reported as mean aligned boundary error in milliseconds. Long-form HMOS uses 200 texts, outputs longer than 60 seconds, and five listeners.

\subsection{Main results}

Relative to Instruction SFT, SCIC-SFT improves Pitch, Energy, Speed, and Pause by 7.31, 10.53, 0.51, and 0.67 percentage points, respectively, with the largest gains on the explicitly conditioned Pitch and Energy attributes. Multi-reward post-training extends the corresponding SCIC-GDPO gains to 9.65, 11.30, 2.97, and 3.87 percentage points. CER remains within a narrow 1.63--1.68\% range, while speaker similarity remains essentially unchanged.

Pause ACC measures whether the detected silence falls inside the requested duration interval. SCIC-GDPO achieves the best result at all three levels, and the mean detected durations of all systems lie inside their corresponding target intervals.

On 200 long-form texts, HMOS increases from 3.244 for Zero-Shot and 3.504 for SFT without instructions to 4.042 with paragraph-level instructions, indicating a clearer expressive hierarchy for the complete system.

The control-magnitude results further show that the control-accuracy gains correspond to stronger direction-normalized changes in each attribute's native domain. Relative to SCIC-SFT, SCIC-GDPO increases the median change magnitude for both directions of Pitch, Energy, and Speed while remaining below the frozen training references.

\subsection{Temporal and codebook-axis ablations}

\begin{table}[t]
\caption{Ablations: control-direction accuracy (\%) and Router Boundary MAE.}
\label{tab:ablation}
\centering
\small
\setlength{\tabcolsep}{5pt}
\begin{tabular}{@{}lrrr@{}}
\toprule
Variant & Pitch & Energy & MAE$\downarrow$ \\
\midrule
SCIC-SFT & 90.14 & 96.39 & 295 ms \\
w/ Hard Router & 88.41 & 96.44 & 375 ms \\
w/o Tag-Specific Codebook Weighting & 89.43 & 94.90 & 292 ms \\
\bottomrule
\end{tabular}
\vspace{-4pt}
\end{table}

Replacing the soft Router with a binary Router reduces Pitch accuracy and increases Boundary MAE from 295 to 375 ms, supporting graded temporal activation. Removing Tag-Specific Codebook Weighting lowers Pitch and Energy accuracy, while its 292-ms Boundary MAE remains close to SCIC-SFT. These results indicate that the Temporal Instruction Router primarily controls instruction scope, whereas Tag-Specific Codebook Weighting mainly controls attribute allocation across residual predictions.

\section{Conclusions}

We introduced SCIC for speaker-relative inline prosody control. RVQ Codebook Diagnosis shows that Energy concentrates in early residual codebooks whereas Pitch requires a deeper prefix, motivating temporal routing and tag-specific codebook weighting. Multi-reward GDPO improves Pitch, Energy, Speed, and Pause control while maintaining comparable intelligibility and speaker similarity. Long-form listening indicates a clearer paragraph-level expressive hierarchy for the complete instruction-conditioned system.


\clearpage
\begin{thebibliography}{25}

\bibitem{instructtts}
Dongchao Yang, Songxiang Liu, Rongjie Huang, Chao Weng, and Helen Meng, ``InstructTTS: Modelling expressive TTS in discrete latent space with natural language style prompt,'' \emph{IEEE/ACM Trans. Audio, Speech, Language Process.}, vol. 32, pp. 2913--2925, 2024.

\bibitem{prompttts}
Zhifang Guo, Yichong Leng, Yihan Wu, Sheng Zhao, and Xu Tan, ``PromptTTS: Controllable text-to-speech with text descriptions,'' in \emph{Proc. IEEE ICASSP}, pp. 1--5, 2023.

\bibitem{prompttts2}
Yichong Leng, Zhifang Guo, Kai Shen, Xu Tan, Zeqian Ju, Yanqing Liu, Yufei Liu, Dongchao Yang, Leying Zhang, Kaitao Song, \emph{et al.}, ``PromptTTS 2: Describing and generating voices with text prompt,'' arXiv preprint arXiv:2309.02285, 2023.

\bibitem{promptstyle}
Guanghou Liu, Yongmao Zhang, Yi Lei, Yunlin Chen, Rui Wang, Zhifei Li, and Lei Xie, ``PromptStyle: Controllable style transfer for text-to-speech with natural language descriptions,'' in \emph{Proc. Interspeech}, pp. 4888--4892, 2023.

\bibitem{controlspeech}
Shengpeng Ji, Qian Chen, Wen Wang, Jialong Zuo, Minghui Fang, Ziyue Jiang, Hai Huang, Zehan Wang, Xize Cheng, Siqi Zheng, \emph{et al.}, ``ControlSpeech: Towards simultaneous zero-shot speaker cloning and zero-shot language style control with decoupled codec,'' arXiv preprint arXiv:2406.01205, 2024.

\bibitem{flespeech}
Hanzhao Li, Yuke Li, Xinsheng Wang, Jingbin Hu, Qicong Xie, Shan Yang, and Lei Xie, ``FleSpeech: Flexibly controllable speech generation with various prompts,'' arXiv preprint arXiv:2501.04644, 2025.

\bibitem{ovinstructtts}
Yong Ren, Jiangyan Yi, Jianhua Tao, Haiyang Sun, Zhengqi Wen, Hao Gu, Le Xu, and Ye Bai, ``OV-InstructTTS: Towards open-vocabulary instruct text-to-speech,'' arXiv preprint arXiv:2601.01459, 2026.

\bibitem{flexivoice}
Dekun Chen, Xueyao Zhang, Yuancheng Wang, Kenan Dai, Li Ma, and Zhizheng Wu, ``FlexiVoice: Enabling flexible style control in zero-shot TTS with natural language instructions,'' arXiv preprint arXiv:2601.04656, 2026.

\bibitem{bagpipertts}
Jinchuan Tian, Haoran Wang, Siddhant Arora, Takashi Maekaku, Keita Goto, Jin Sakuma, Yusuke Shinohara, Chao-Han Huck Yang, and Shinji Watanabe, ``Bagpiper-TTS: Natural language guided universal speech synthesis,'' arXiv preprint arXiv:2606.22811, 2026.

\bibitem{cosyvoice3}
Zhihao Du, Changfeng Gao, Yuxuan Wang, Fan Yu, Tianyu Zhao, Hao Wang, Xiang Lv, Hui Wang, Chongjia Ni, Xian Shi, \emph{et al.}, ``CosyVoice 3: Towards in-the-wild speech generation via scaling-up and post-training,'' arXiv preprint arXiv:2505.17589, 2025.

\bibitem{hdppt}
Sihang Nie, Xiaofen Xing, Jingyuan Xing, Baiji Liu, and Xiangmin Xu, ``HD-PPT: Hierarchical decoding of content- and prompt-preference tokens for instruction-based TTS,'' in \emph{Proc. IEEE ICASSP}, pp. 16487--16491, 2026.

\bibitem{sparktts}
Xinsheng Wang, Mingqi Jiang, Ziyang Ma, Ziyu Zhang, Songxiang Liu, Linqin Li, Zheng Liang, Qixi Zheng, Rui Wang, Xiaoqin Feng, \emph{et al.}, ``Spark-TTS: An efficient LLM-based text-to-speech model with single-stream decoupled speech tokens,'' arXiv preprint arXiv:2503.01710, 2025.

\newpage
\bibitem{voxtream2}
Nikita Torgashov, Gustav Eje Henter, and Gabriel Skantze, ``VoXtream2: Full-stream TTS with dynamic speaking rate control,'' arXiv preprint arXiv:2603.13518, 2026.

\bibitem{restyle}
Haitao Li, Chunxiang Jin, Chenglin Li, Wenhao Guan, Zhengxing Huang, and Xie Chen, ``ReStyle-TTS: Relative and continuous style control for zero-shot speech synthesis,'' arXiv preprint arXiv:2601.03632, 2026.

\bibitem{magictts}
Jialong Mai, Xiaofen Xing, and Xiangmin Xu, ``MAGIC-TTS: Fine-grained controllable speech synthesis with explicit local duration and pause control,'' arXiv preprint arXiv:2604.21164, 2026.

\bibitem{wordvoice}
Sihang Nie, Jinxin Ji, Xiaofen Xing, Deyi Tuo, Chengbin Jin, Jialong Mai, and Xiangmin Xu, ``WordVoice: Explicit and decoupled multi-dimensional word-level control for LLM-based TTS,'' arXiv preprint arXiv:2607.06461, 2026.

\bibitem{qwenaudio3tts}
Bajian Xiang, Cheng Wen, Han Zhao, Hao Wang, Haoxu Wang, Jiawei Jin, Jiayan Cui, Jie Chen, Mengxi Nie, Tianyu Zhao, \emph{et al.}, ``Qwen-Audio-3.0-TTS: Freely controllable and highly robust speech synthesis with multi-stage training paradigm,'' arXiv preprint arXiv:2607.23938, 2026.

\bibitem{fishaudios2}
Shijia Liao, Yuxuan Wang, Songting Liu, Yifan Cheng, Ruoyi Zhang, Tianyu Li, Shidong Li, Yisheng Zheng, Xingwei Liu, Qingzheng Wang, \emph{et al.}, ``Fish Audio S2 technical report,'' arXiv preprint arXiv:2603.08823, 2026.

\bibitem{any2speech}
Xingchen Song, Di Wu, Dinghao Zhou, Pengyu Cheng, Hongwu Ding, Yunchao He, Jie Wang, Shuai Wang, Shengfan Shen, Sixiang Lv, \emph{et al.}, ``Any2Speech: Borderless long audio synthesis,'' arXiv preprint arXiv:2603.19798, 2026.

\bibitem{emotratts}
Tianchi Liu, Zeyang Song, Tianrui Wang, Zhipeng Li, Chenglin Xu, and Yiwen Guo, ``EmoTra-TTS: Smooth intra-utterance emotion transitions for speech synthesis,'' arXiv preprint arXiv:2608.23791, 2026.

\bibitem{qwen3tts}
Hangrui Hu, Xinfa Zhu, Ting He, Dake Guo, Bin Zhang, Xiong Wang, Zhifang Guo, Ziyue Jiang, Hongkun Hao, Zishan Guo, \emph{et al.}, ``Qwen3-TTS technical report,'' arXiv preprint arXiv:2601.15621, 2026.

\bibitem{gdpo}
Shih-Yang Liu, Xin Dong, Ximing Lu, Shizhe Diao, Peter Belcak, Mingjie Liu, Min-Hung Chen, Hongxu Yin, Yu-Chiang Frank Wang, Kwang-Ting Cheng, \emph{et al.}, ``GDPO: Group reward-decoupled normalization policy optimization for multi-reward RL optimization,'' arXiv preprint arXiv:2601.05242, 2026.

\bibitem{lora}
Edward J. Hu, Yelong Shen, Phillip Wallis, Zeyuan Allen-Zhu, Yuanzhi Li, Shean Wang, Lu Wang, and Weizhu Chen, ``LoRA: Low-rank adaptation of large language models,'' in \emph{Proc. ICLR}, 2022.

\bibitem{qwen3asr}
Xian Shi, Xiong Wang, Zhifang Guo, Yongqi Wang, Pei Zhang, Xinyu Zhang, Zishan Guo, Hongkun Hao, Yu Xi, Baosong Yang, \emph{et al.}, ``Qwen3-ASR technical report,'' arXiv preprint arXiv:2601.21337, 2026.

\bibitem{wespeaker}
Hongji Wang, Chengdong Liang, Shuai Wang, Zhengyang Chen, Binbin Zhang, Xu Xiang, Yanlei Deng, and Yanmin Qian, ``WeSpeaker: A research and production oriented speaker embedding learning toolkit,'' arXiv preprint arXiv:2210.17016, 2022.

\end{thebibliography}
\end{document}